# Atomic-Scale Movement Induced in Nano-Ridges by Scanning Tunneling Microscopy on Epitaxial Graphene Grown on 4H-SiC(0001)

Peng Xu, Steven D. Barber, J. Kevin Schoelz, Matthew L. Ackerman, Dejun Qi, and Paul M. Thibado[a),b)]

Department of Physics, University of Arkansas, Fayetteville, Arkansas 72701

Virginia D. Wheeler, Luke O. Nyakiti, Rachael L. Myers-Ward, C. R. Eddy, Jr., and D. Kurt Gaskill[b)]

U.S. Naval Research Laboratory, Washington, District of Columbia 20375

[a)]American Vacuum Society member.
[b)]Electronic mail: thibado@uark.edu, gaskill@nrl.navy.mil

Nanoscale ridges in epitaxial multilayer graphene grown on the silicon face of 4º off-cut 4H-SiC (0001) were found using scanning tunneling microscopy (STM). These nano-ridges are only 0.1 nm high and 25-50 nm wide, making them much smaller than previously reported ridges. Atomic-resolution STM was performed near and on top of the nano-ridges using a dual scanning technique in which forward and reverse images are simultaneously recorded. An apparent 100% enlarged graphene lattice constant is observed along the leading edge of the image for both directions. Horizontal movement of the graphene, due to both an electrostatic attraction to the STM tip and weak bonding to the substrate, is thought to contribute to the results.

# I. INTRODUCTION

The novel electronic properties of graphene have been inspiring intense research efforts since this two-dimensional material was first successfully isolated in 2004.[1] Its intriguing features, such as ballistic transport,[2] the quantum Hall effect,[3] and ultra-high mobility,[4] mark graphene as a potentially pivotal material in the field of carbon-based electronics. In particular, epitaxial graphene grown on SiC has been identified as one of the most likely avenues to graphene-based electronics.[5, 6] Semi-insulating SiC has a large band gap which makes it ideally suited for the production of top-gated, electrically isolated devices, and furthermore, it is already available in the form of large-diameter wafers compatible with current industrial technology.[7, 8] To form graphene on SiC,[9] the sample must be annealed at temperatures >1200°C, causing Si to sublimate and leaving behind large regions of pristine epitaxial graphene once the sample has cooled.[10-12] Interestingly, atomic force microscopy (AFM) images on the graphene surface have revealed distinctive, large ridges (~1.5 nm high) running along step edges and across terraces, with atomically flat regions in between.[13-16] The appearance of these ridges has been attributed to strain-induced bending and buckling of the graphene caused by the sign difference in the thermal expansion coefficients of graphene and SiC as the sample cools.[17-19] Scanning tunneling microscopy (STM) studies have confirmed this interpretation and have shown that the large ridges can be manipulated and even generated through interactions with the STM tip during imaging.[20]

Scanning probe microscopy techniques have also been helpful toward gaining insight into the crucial electronic interaction between the graphene surface layer and the SiC substrate.[21] For instance, an STM study by Rutter *et al.* was responsible for determining that the dominant source of electron scattering in epitaxially grown graphene was due to defects in the SiC substrate.[22] And to help reduce such detrimental doping effects, a later STM study found that the carbon buffer layer[23] at the graphene/SiC interface could be decoupled from the SiC substrate by means of fluorine intercalation.[24] In a separate study, Yakes *et al.* used a four-probe STM system to perform local transport measurements and discovered a local conductance anisotropy in the epitaxial graphene.[25] The direction of increased scattering was attributed to interference from Si atoms trapped underneath the graphene along the step edges of the SiC crystal where Si desorption occurs more rapidly.[26-28] These small-scale electronic studies and others have proven invaluable to uncovering the fundamental interactions governing transport in epitaxial graphene, and the same could be said for investigations into large-scale topographic features like the aforementioned ridges and terrace steps. What remains largely unexamined, however, is the very small-scale topography in epitaxial graphene on SiC and its response to local imaging.

In this work, we report nano-ridges (~0.1 nm high) found by an STM study of the multilayer graphene on 4º off-cut 4H-SiC(0001) (the Si face). These ridges are much smaller than those found in earlier studies, but they likely form by the same mechanism (i.e., thermal coefficient of expansion mismatch). Surprisingly, images acquired near the nano-ridges show an enlarged graphene lattice constant on the left or right edge depending on the scanning direction. Moreover, increasing the tunneling current setpoint increased the size of the altered region in the STM image. We propose that an electrostatic attractive force between the STM tip and graphene sample produces these results by dynamically displacing the surface layer.

## II. EXPERIMENTAL

The epitaxial graphene sample used in this study was grown on 20 μm of intentionally n-doped ($1\times10^{14}$/cm$^3$) epitaxial 4H-SiC layer on the Si face of a 4º off-cut 4H-SiC substrate measuring 16 mm × 16 mm and cut from a 76.2 mm diameter parent wafer (Cree, Inc). Growth was carried out in a commercially available hot-wall Aixtron VP508 chemical vapor deposition reactor. Prior to graphene growth, the substrate was etched *in situ* in a H$_2$ ambient environment for 5 minutes at 1520 °C. This etching produces a controlled starting surface that is dominated by SiC surface steps roughly 0.5 nm high. After the H$_2$ etching step, the ambient environment was switched to Ar with a transition period of 2 minutes during which pressures varied by ±50% around 100 mbar. The subsequent 120 minute graphene growth process was conducted under a flowing Ar environment of 20 standard liters per minute at 100 mbar, with a substrate growth temperature of 1620 °C.[29] After growth, the sample was cooled to room temperature, cut to 7 mm × 14 mm, and diamond scribed labels were added to the carbon face. Next, Raman data was collected for the sample using a Thermo DXR system. A 532 nm, 8 mW laser was used as the pump probe with a spot size 0.6 μm. After characterization with Raman, the sample was secured, sealed, and sent to the STM facility.

Experimental STM images were obtained using an Omicron ultrahigh-vacuum (base pressure is 10$^{-10}$ mbar) STM (low-temperature model) operated at room temperature. The sample was mounted with silver paint onto a flat tantalum sample plate and transferred through a load-lock into the STM chamber where it was electrically grounded. No cleaning or heating was done to the sample. STM tips were electrochemically etched from 0.25 mm diameter tungsten wire via a custom double lamella setup with an automatic gravity-switch cutoff.[30] After etching, they were gently rinsed with distilled water, briefly dipped in a concentrated hydrofluoric acid solution to remove surface oxides,[31] and then loaded into the STM chamber through the same load-lock. All STM images were acquired using a positive tip bias of 0.1 V.

## III. RESULTS AND DISCUSSION

### A. Raman Spectra and STM Images of Nano-ridges

A full-range Raman scan of the epitaxial graphene on the SiC sample is shown in Fig. 1. The largest peaks are identified with the SiC substrate; in addition both the G and 2D peaks of graphene are present. A magnified view of the 2D peak is shown in the inset and has a full width at half maximum of 70 cm$^{-1}$, which indicates that there are over 2 layers of graphene on the surface of our sample.[32]

Two large-scale, filled-state STM images of the epitaxial graphene on the SiC sample obtained with a tunneling current setpoint of 0.05 nA are shown in Fig. 2. The first image, displayed in Fig. 2(a), is 300 nm × 300 nm and exhibits two prominent features. First, the sharp contrast between the dark upper half and light lower half indicates the presence of a step in the SiC substrate. A height data line profile extracted from the image is plotted below it (the location is indicated by the black line drawn on the image) and reveals that the step is about 0.55 nm high (note, the c lattice constant for 4H-SiC is 1.0053 nm, which is 4 SiC bilayers). Second, nano-ridges are seen running from top to bottom which are uninterrupted by the presence of the step. Some appear to have distinct kinks and some bunch together, yet without combining to form wider ridges. A slightly closer view of the nano-ridges is provided in the 200 nm × 200 nm image of Fig. 2(b). A height profile was again extracted, this time horizontally, and is displayed below the image. It shows that, on average, the nano-ridges are around 0.1 nm high and between 25-50 nm wide. They exhibit an oscillatory nature, and though not readily apparent in the image, smaller peaks sometimes occur between the more prominent ones.

Previously observed large ridges mentioned earlier are much higher than the nano-ridges shown here in Fig. 2. Scanning probe experiments report that the larger ridges are 1 to 2 nm high (10 to 20 times larger).[14] In addition to the nano-ridges in Fig. 2 being much smaller, they also have a very different geometry. They tend to meander, have kinks, and bunch together, while the large ridges tend to be isolated and straighter. The kinks are especially interesting because they may represent nucleation or pinning sites for the formation of the nano-ridges. Nevertheless, it seems likely that both types of ridges are formed by the same physical process, namely, compressive strain[33] induced during cooling of the SiC wafer.[17, 19, 34] As the system is cooled from the growth temperature, the graphene expands as the substrate contracts,[35] and the resulting strain is most easily relieved by having portions of the graphene flex away from the substrate and buckle. The nano-ridges discussed here are likely the source for the extra graphene material required to form the larger ridges found in other studies with AFM. Other studies may have missed the nano-ridges due to their small, atomic-scale heights or possibly due to the stronger interaction of AFM.[13]

## B.   Atomic Scale Movement Induces by STM in Nano-ridges

Atomic-scale STM was performed in the region of the nano-ridges and two typical 12 nm × 12 nm filled-state images are shown in Fig. 3(a,b). These images were acquired simultaneously at the same location using a dual scanning technique, a tunneling current of 0.5 nA, and the fast-scan direction horizontal with the tip moving at a speed of 72 nm/s. The image recorded while the STM tip moved from left to right on each line of data (with 400 lines of data) is displayed in Fig. 3(a). The hexagonal symmetry structure of graphene is resolved in various smaller regions throughout and indicating the presence of graphene on the top of the sample. However, the lattice constant of the graphene appears to change as the image is viewed from left to right. In fact, within approximately 0.5 nm of the entire leftmost edge, the lattice constant is enlarged by 150% horizontally and 100% vertically when compared with the right side. On the right-hand side, the lattice constant is measured to be 0.25 nm, consistent with theory and previous experiments.[36]

STM data was also simultaneously recorded as the tip returned moving right to left for each line of data and this is shown in Fig. 3(b). Note that this data acquisition process does not result in mirror images. Rather, the position marked by the ovals represents the same location on the sample surface in both instances. This second image is nearly identical to the previous one; however, there is one important difference. The enlarged lattice constant is now on the right edge while the normal lattice constant is on the left side. Because the enlarged lattice constant is seen on the left in Fig. 3(a) and on the right in Fig. 3(b), the possibility of it being a real permanent feature in the graphene lattice is automatically ruled out. It is also not a result of in-plane stretching of the carbon-carbon bonds, due to their stiffness as well as the relatively low strain of the epitaxial graphene (~1%).[37]

## C. Discussion

The stretched out atomic features in these images are more reminiscent of data containing piezoelectric creep.[38] When an STM image is acquired during the creep time, the beginning part of the image will appear warped or sheered compared to the ending part. Creep results are not real, but simply an artifact of the tip moving more than the assigned value. However, creep quickly goes away with typical scan speeds, yet the effect shown in Fig. 3(a,b) never goes away. Even though the lack of time dependence in our data successfully rules out the piezoelectric creep model, we do believe that independent movement between the tip and sample is occurring outside the assigned data acquisition values. Piezoelectric hysteresis is another mechanism that can lead to distorted STM images; however, this effect does not change in time. For this case, the voltage applied to the piezoelectric scanner leads to a non-linear response in its displacement.[39-41] For our studies, the piezoelectric scanner is being operated within its linear range. Also, this type of distortion does not occur for our control samples (not shown) like graphite and graphene on copper.

Previous studies have established that graphite derivatives, including graphene, are especially susceptible to movement induced by the electrostatic interaction between the STM tip and the sample.[42-45] In the simplest situation, constant-current scanning tunneling spectroscopy measurements used on freestanding graphene clearly demonstrated that the graphene flexes more than 30 nm toward the STM as the tip bias increases from 0.1 V to 3.0 V.[43] Using a similar technique on graphite, large scale movement of 20 nm wide ribbons was also demonstrated.[44] Another important example, on an insulating substrate, was done by Mashoff et.al, who demonstrated oscillatory motion induced in graphene on SiO$_2$ using STM.[46]

With the idea of independent sample movement in mind, we introduce a "sample creep" model to best explain our results. In this model, as the STM tip moves across the graphene surface, the top graphene layer will be attracted to the tip and slide horizontally with the tip, which causes the observed features. This is especially plausible given the weak coupling between graphene and the substrate below, as well as the excess graphene available on the surface in the form of nano-ridges. A series of four STM schematics is provided at the bottom of Fig. 3 to illustrate how sample creep would replicate the unusual details in the findings. Each diagram portrays, from top to bottom, a moving triangular STM tip and the path it traces (labeled STM), the semi-mobile graphene surface of constant local density of states (labeled G), and the stationary substrate (labeled Sub.). As the tip first begins to move left to right, corresponding to the left side of Fig. 3(c), it drags the graphene along with it (motion direction indicated by the arrow), though at a slower speed than the tip is moving. The relative motion between the two causes the tip to trace a path with a larger lattice constant than if the sample were stationary. Soon, however, the available slack in the graphene has been removed, or alternatively, a compressive strain has accumulated ahead of the tip. At this point, the graphene ceases to slide, and it is imaged with its normal lattice constant, as depicted in the STM trace shown on the right side of Fig. 3(c). When the tip reaches the end of a given line, and changes direction, this allows the graphene surface to relax back to equilibrium and, in addition, allows the tip to now pull the graphene in the opposite direction. Similar to before, the motion of the graphene (indicated by the arrow) is in the same direction as the STM tip and falsely constructs an enlarged lattice constant, as illustrated in the right side of Fig. 3(d). At some point the motion of the graphene is constrained, and the rest of the line is imaged normally. This corresponds with the left side of Fig. 3(d). As the process just described repeats for every line, it results in the time-independent edge distortions observed line-by-line going up the images. Thus instead of unrecorded movement of the piezoelectric causing these deformations, we are seeing movement of the sample surface caused by both its electrostatic attraction to the STM tip and weak bonding to the underlying layer(s).

We can further test our model by examining the width of the distorted area as a function of the setpoint current, since piezoelectric creep is independent of the tunneling current. To do this, we acquired a large set of STM images all from the same location (but at a different one than in Fig. 3). The constant-current STM images were collected at various setpoints between 0.02 nA and 2 nA, and all with a common scan speed of 72 nm/s. The current setpoint does have a clear effect on the edge width, as seen in the semi-log plot shown in Fig. 4(a). As the tunneling current was increased, the edge width approximately doubled from 0.5 nm to more than 1 nm. This increase in width can be explained by the sample creep model, since increasing the current setpoint effectively shortens the distance between the tip and sample and therefore increases the electrostatic attraction between them. As a result, the tip pulls on the sample with a greater force, resulting in greater displacements of the graphene and larger edge widths.

The edge width was also studied as a function of the STM tip scan speed for a fixed setpoint current of 2 nA as shown in Fig. 4(b) and overlaid with a linear regression of the data. The scan speed was varied between 50 and 150 nm/s, and over this range the edge width generally increased with scan speed, from 0.75 nm to 1.5 nm. This trend is also consistent with the sample creep model, but also lends some insight into the mechanism. At the end of each scan line where the STM tip reverses direction, the accumulated strain in the graphene is released as the tip begins to pull it in the opposite direction. If the tip is moving very slowly, then the graphene will relax prior to the STM tip imaging the surface. This would result in measuring the smaller edge width observed experimentally.

The horizontal shifting of the graphene sample surface agrees favorably with interpreting the nano-ridges as buckled graphene which has separated from the underlying layers. Since the nano-ridges are physically separated from the surface, the interaction with the substrate is significantly reduced. One would then expect the nano-ridges to act as superior conductance channels. This is because it was found that as graphene becomes partially decoupled from a SiC substrate, electron scattering will be reduced.[22, 24] A similar outcome was reported by Pirkle *et al.* when studying the electrical properties of chemical vapor deposited graphene transferred onto $SiO_2$.[47] In particular, they found via AFM measurements across graphene strips that the closer the graphene was to the substrate the lower the electron mobility.[48] Raised nano-ridges could, therefore, also lead to conductance anisotropy, just as step edges do on a larger scale.[25] Since the ridges occur in essentially parallel rows, local transport would favor the direction following the ridges rather than crossing them.

## IV. CONCLUSIONS

In conclusion, we have performed an atomic scale STM study on an atomically flat region of multilayer epitaxial graphene grown on 4º off-cut 4H-SiC(0001). For the first time, atomic-scale parallel nano-ridges were observed on the graphene layer. Atomic-resolution STM images were obtained near and on top of these ridges and significantly enlarged lattice constants were investigated. We showed that the graphene top layer, during scanning, moves with the STM tip due to an electrostatic attraction between the tip and sample, which is consistent with the idea that the ridges are regions of the graphene sheet that have buckled due to strain and have become partially decoupled from the underlying layer.

## ACKNOWLEDGMENTS

P. X. and P. T. gratefully acknowledge the financial support of the Office of Naval Research under grant number N00014-10-1-0181 and the National Science Foundation under grant number DMR-0855358. Work at the U.S. Naval Research Laboratory is supported by the Office of Naval Research. LON gratefully acknowledges postdoctoral fellowship support through the ASEE.

**Figure Captions**

FIG. 1. Raman spectra of graphene on SiC(0001) obtained using 532 nm laser line. Inset: detailed feature of 2D band for graphene.

FIG. 2. (Color online) Filled-state STM images of graphene on SiC(0001) taken with a current setpoint of 0.05 nA and tip bias of 0.1 V. (a) 300 nm × 300 nm image showing parallel nano-ridges prominently aligned vertically across a 0.55 nm horizontal step. A height line profile extracted from the image showing the step height is displayed below the image. (b) 200 nm × 200 nm image shows a zoomed in view of the nano-ridges. A height line profile extracted from the image is displayed below.

FIG. 3. (Color online) Filled-state 12 nm × 12 nm STM images of graphene grown on SiC(0001) taken in an area with nano-ridges using a current setpoint of 0.5 nA and a tip bias of 0.1 V. Both images were acquired simultaneously from the same location using a dual scanning technique which recorded height data when the STM tip was moving both (a) left to right and (b) right to left. Illustrations of the semi-mobile sample and the contour recorded by the STM tip as it scans from (c) left to right and (d) right to left.

FIG. 4. The width of the area containing an enlarged lattice constant measured from atomic-resolution STM images of graphene on SiC as functions of (a) current setpoint, but with a common scan speed of 72 nm/s and (b) scan speed, but with a common current setpoint of 2 nA.

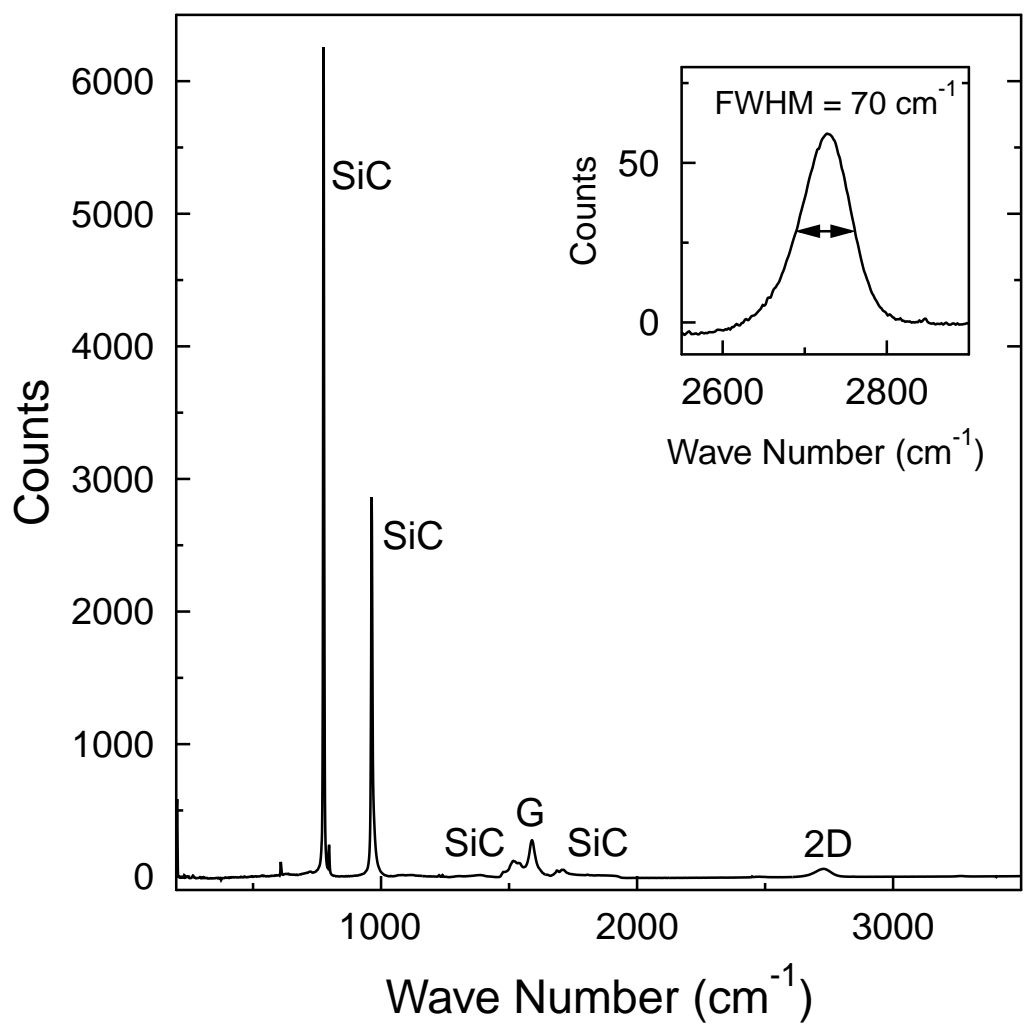

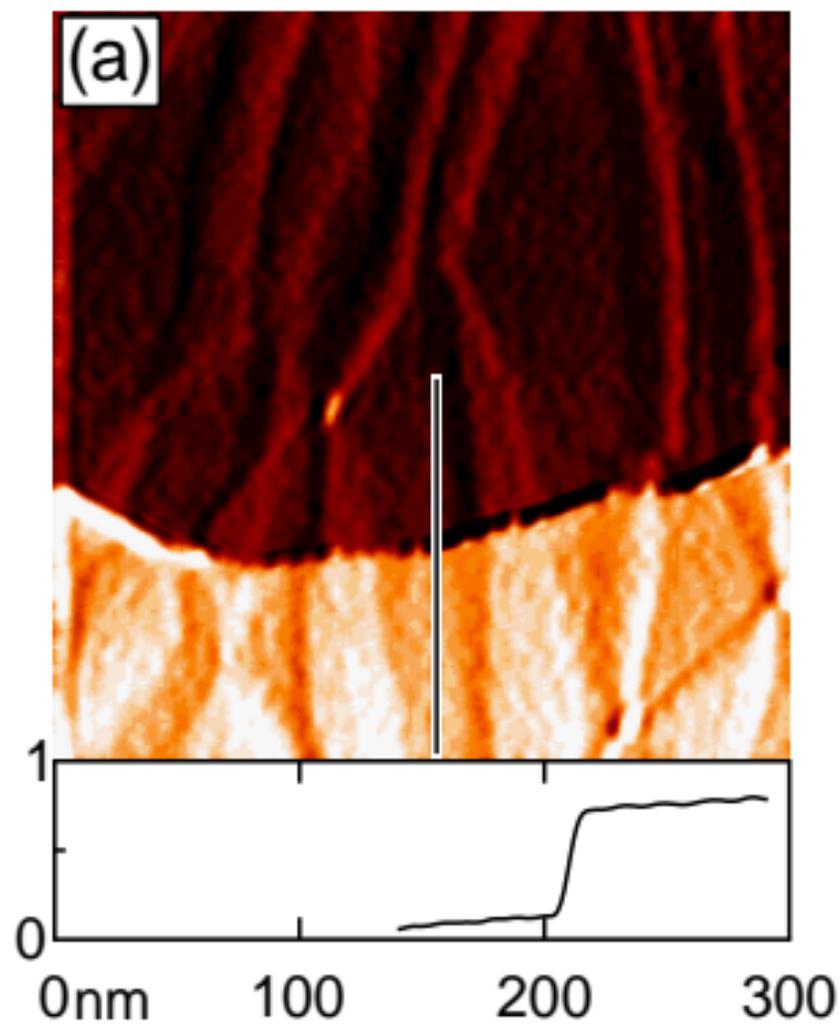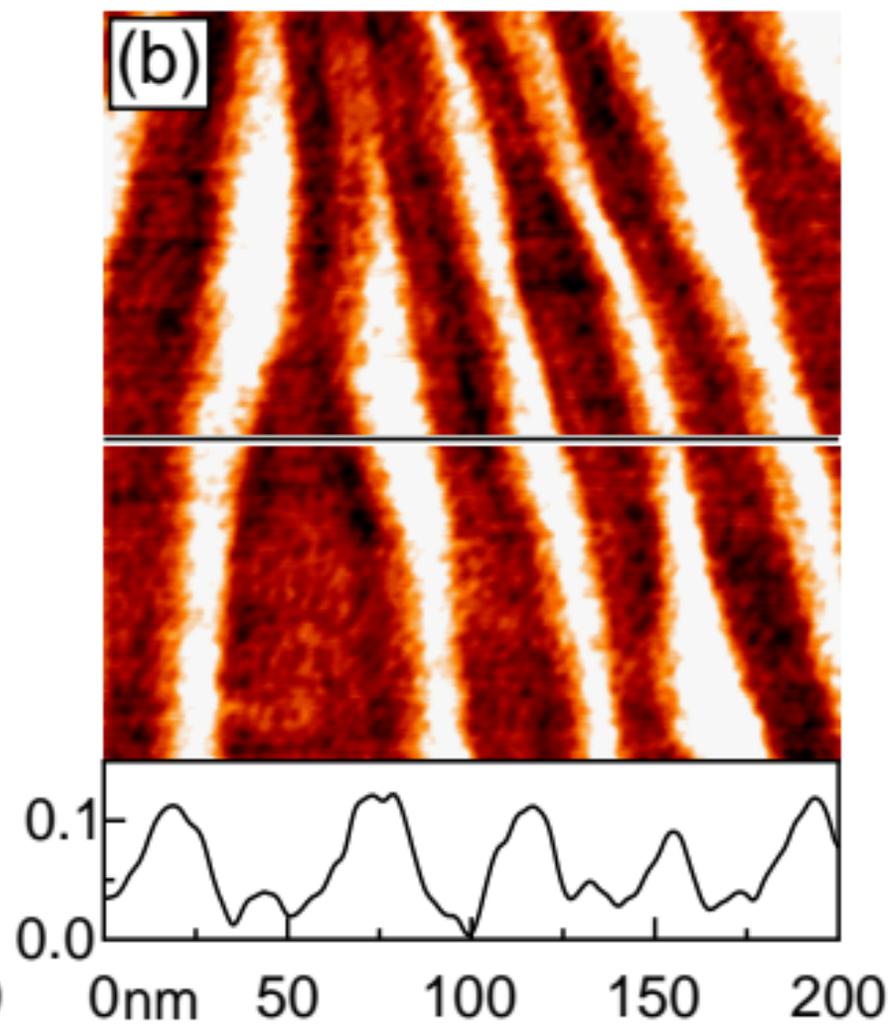

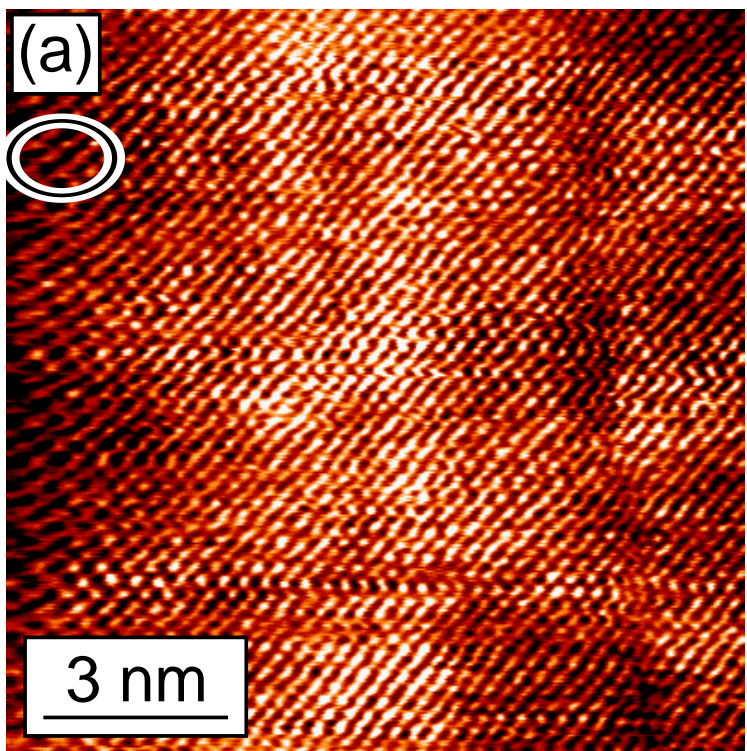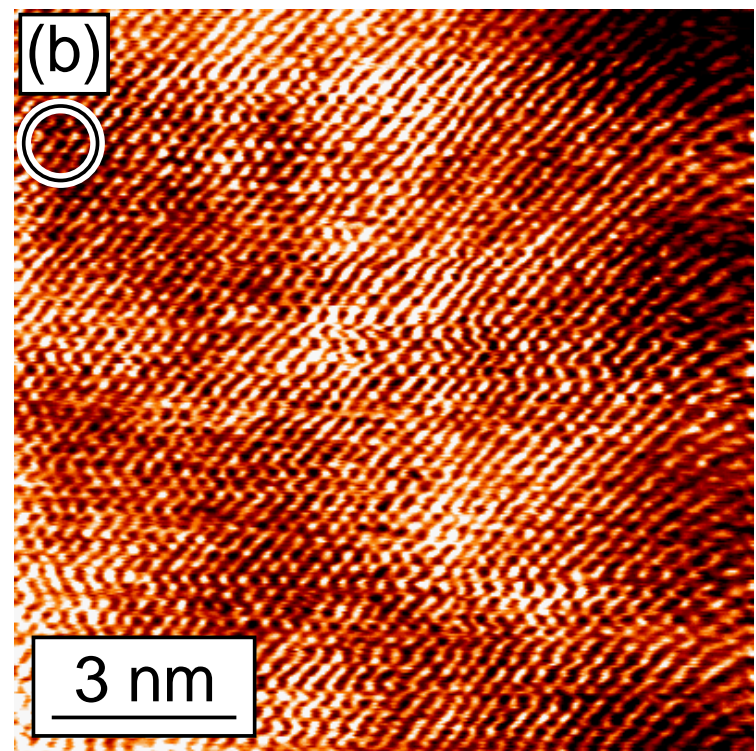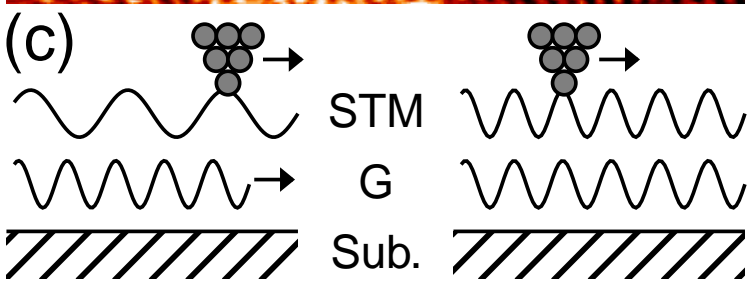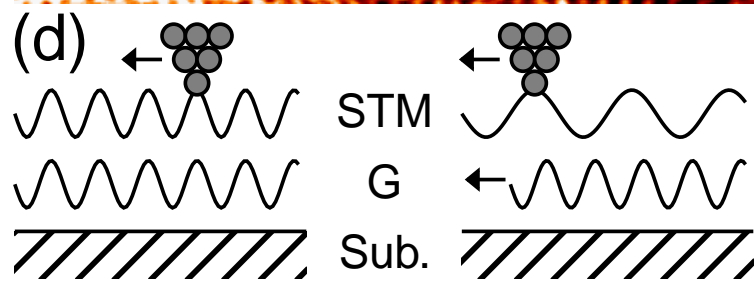

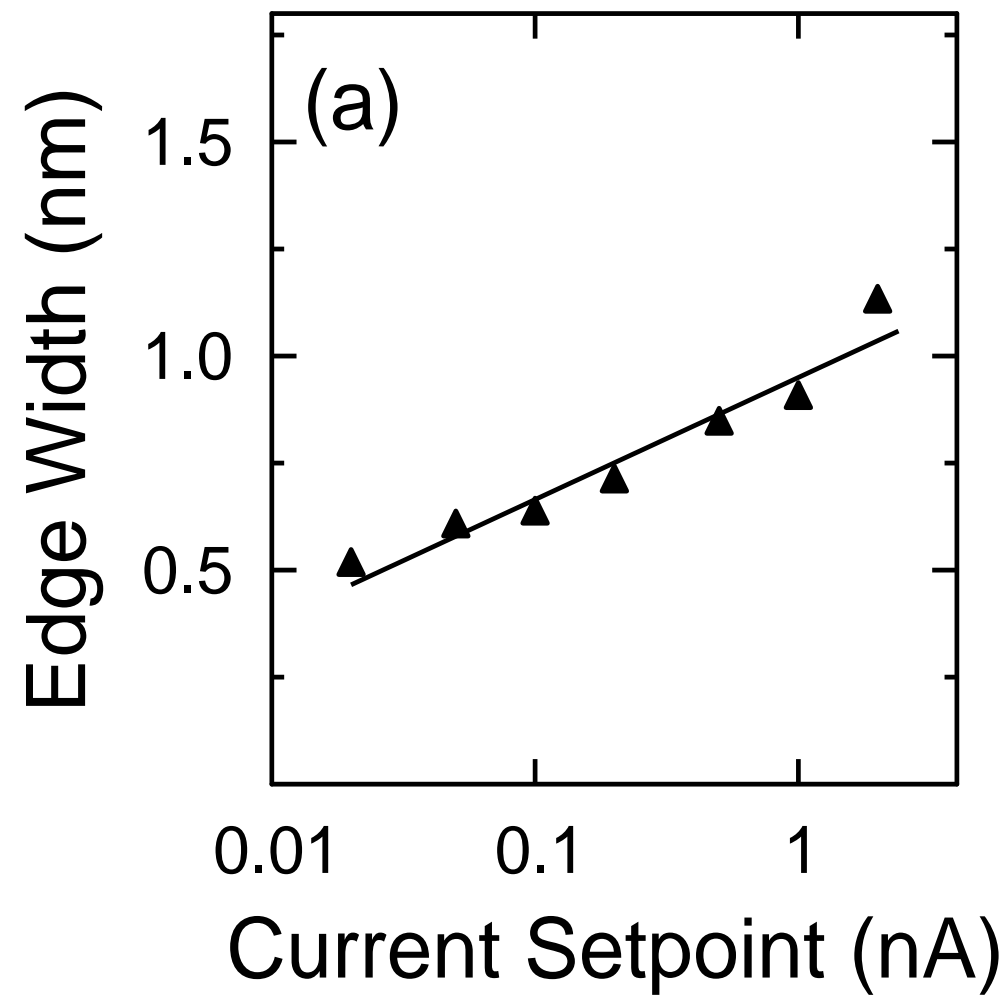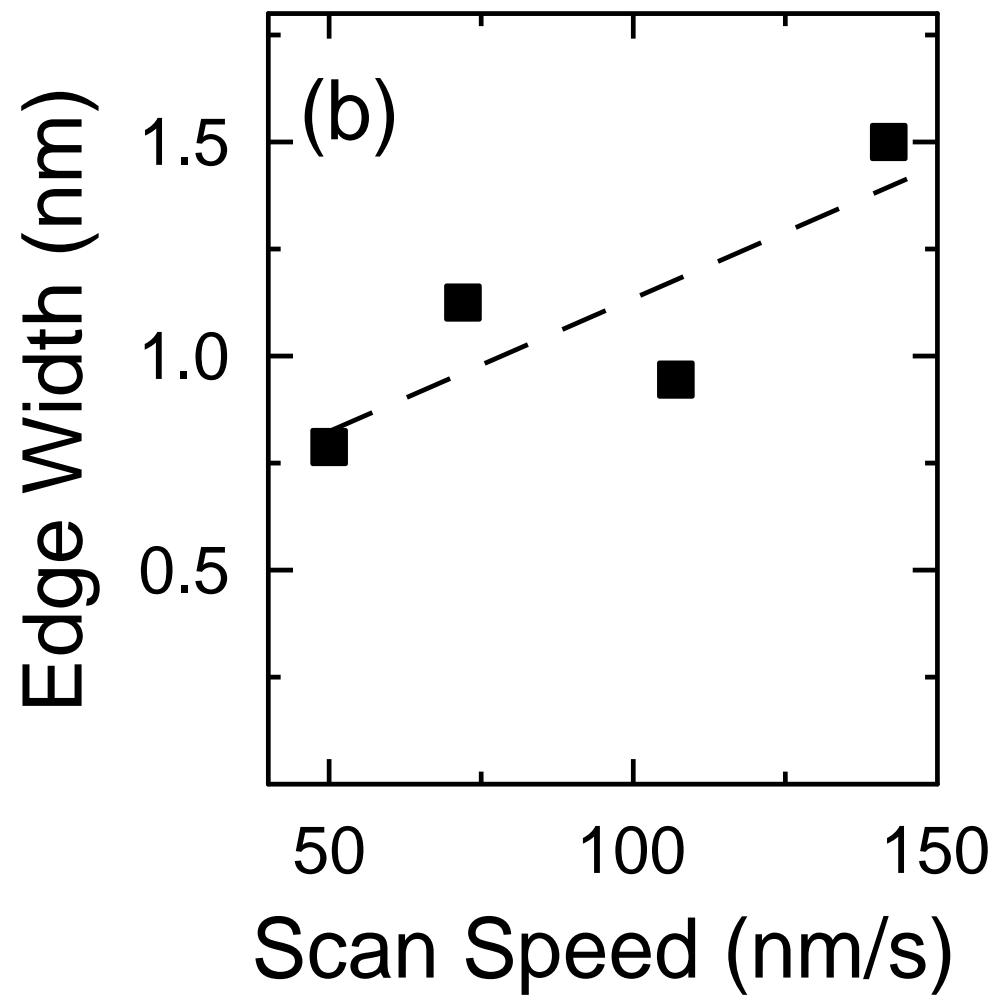